\documentclass{article}
\usepackage{spconf,amsmath,graphicx}
\usepackage{url}

\title{Progressive Voice Trigger Detection: Accuracy vs Latency}
\name{Siddharth Sigtia*\thanks{* Equal contribution}, John Bridle*, Hywel Richards, Pascal Clark, Erik Marchi, Vineet Garg}
\address{Apple}
\begin{document}
%
\maketitle
\begin{abstract}

We present an architecture for voice trigger detection for virtual assistants. The main idea in this work is to exploit information in words that immediately follow the trigger phrase. 
We first demonstrate that by including more audio context after a detected trigger phrase, we can indeed get a more accurate decision. 
However, waiting to listen to more audio each time incurs a latency increase. \emph{Progressive} Voice Trigger Detection allows us to trade-off latency and accuracy by accepting clear trigger candidates quickly, but waiting for more context to decide whether to accept more marginal examples. 
Using a two-stage architecture, we show that by delaying the decision for just 3\% of detected true triggers in the test set, we are able to obtain a relative improvement of 66\% in false rejection rate, while incurring only a negligible increase in latency. 


\end{abstract}
\section{Introduction}
\label{sec:intro}

Virtual assistants allow users to interact with devices such as speakers, phones, watches and headphones via voice commands. Typically, voice commands from a user are prefixed with a \emph{trigger phrase}. The presence of the trigger phrase at the beginning of an utterance helps distinguish audio that is directed towards the assistant from background speech. The problem of accurately detecting a trigger phrase is known as voice trigger detection \cite{MLBlogHS,bridle1973efficient}, wake-word detection \cite{kumatani2017direct,jose2020accurate}, or keyword spotting \cite{266505,fernandez2007application,chen2014small,lin2020training,rose1990hidden}. 

This work is motivated by the observation that audio following the trigger phrase can contain a strong signal about whether an utterance was directed towards the assistant or not. 
We found that the top 10 most popular words that follow the trigger phrase, \textbf{account for 80\% of the distribution}. 
While the top 20 words account for 90\% of the distribution. 
Although it is clear that audio following the trigger phrase can help in determining whether an utterance is directed towards the assistant, this improved accuracy comes at the cost of latency. Therefore, to design a practical voice trigger detector that runs on-device, simply waiting to listen to more audio is not sufficient.

\begin{figure}[t]
\begin{minipage}[]{1\linewidth}
  \centering
  \centerline{\includegraphics[width=0.8\textwidth]{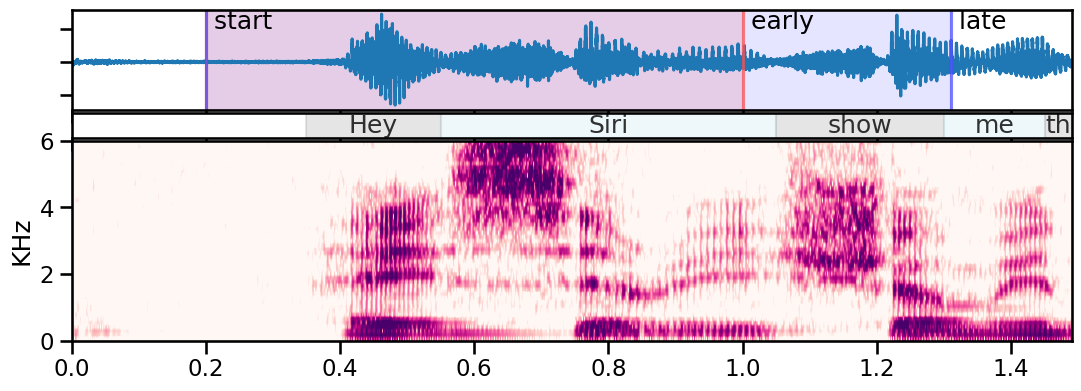}}
  \caption{Two-stage design: An example of early and late decision boundaries used in this paper. Early decisions result in a faster response, but later decisions provide greater accuracy.}
  \label{windows}
\end{minipage}
\end{figure}

 In this study, we first present a model and experiments that demonstrate that detection accuracy does indeed improve as we add more audio after the trigger phrase, i.e. the \emph{same model} is able to \emph{progressively} improve its estimates as we add more audio context after the trigger phrase. We then devise a two-stage architecture where the model produces an early and a late score (Figure \ref{windows}). We show that the early score is sufficient for a majority of examples in the test set, while the late score allows us to make better decisions for more difficult/marginal cases. This two-stage design allows us to achieve a favourable balance between accuracy and latency. We show that by delaying triggering for only 3\% of true utterances in the test set, we are able to reduce the number of false rejects by 66\% for the same FA rate, while incurring a 17\% relative increase in expected latency for accepted true triggers. Note that a similar idea was recently proposed in \cite{wangaudio,kumarbuilding}, however the model in their work is used to \emph{verify} whether a given segment contains the trigger phrase \emph{on the server}, whereas our proposal is for an on-device voice trigger detector. 

\vspace{-2.5mm}
\section{Model}
\label{sec:format}

We employ a two-stage architecture for voice trigger detection \cite{MLBlogHS}. The first-stage comprises a low-power detector that processes streaming audio and is \emph{always-on} \cite{sigtia2018,higuchi2020s1dcnn}. If a detection is made at the first stage, the detector marks the start and end points 
of the purported keyword segment (Figure \ref{windows}) and the segment is then re-scored by larger, more complex models \cite{sigtia2020mtl,adya2020hybrid}. Note that this paper is concerned only with the larger models in the second-pass.

\subsection{Architecture}


Following previous work \cite{sigtia2020mtl,adya2020hybrid,sigtia2020mtlspk}, the input to the model is a 40-dimensional Mel-filterbank which is computed at a rate of 100 frames-per-second (FPS) from the input audio. We stack 7 contiguous frames together to form an input window and down-sample the sequence of windows by a factor of 3. These features are input to a stack of 4 bidirectional LSTM layers with 256 units in each of the forward and backward layers. The network is trained using multi-task learning (MTL) \cite{caruana1997multitask}, by minimising 2 different objectives simultaneously \cite{sigtia2020mtl,sigtia2020mtlspk}. 

The first objective is to assign the highest score to the correct sequence of context-independent phonetic labels. This is done by minimising the connectionist temporal classification (CTC) loss \cite{ctc_graves}.  The output layer or \emph{head} corresponding to this objective comprises an affine transformation followed by a softmax non-linearity \cite{bridle1990probabilistic} over 54 output units. These units cover the set of context independent  phonemes, word and sentence boundaries and the blank symbol used by the CTC loss. We refer to this loss as the \emph{phonetic loss} in the rest of the paper. 

The second objective used to train the model is a binary sequence classification loss. The positive class corresponds to utterances that are intended for the assistant. These examples are of the type ``\textbf{$\langle$trigger phrase$\rangle$, $\langle$payload$\rangle$}". The negative class corresponds to difficult examples that result in false detections. The output head for this loss contains an affine transformation and a softmax non-linearity over 2 units, the positive and negative classes. The sequence classification loss is defined as follows: 

\begin{equation*}\label{eq:pareto mle2}
\begin{aligned}
C_{pos} = C_{\text{CTC}}(\text{Trigger} | \mathbf{X}), \\
C_{neg} = -\log \sum_t y_t^{n},
\end{aligned}
\end{equation*}
where $C_{\text{CTC}}$ represents the CTC loss function for \emph{positive examples}, i.e. the input features $\mathbf{X}$ contain the trigger phrase, $y_t^{n}$ represents the network output for the negative class at time $t$, $C_{pos}$, $C_{neg}$ represent the losses for positive and negative classes respectively. We refer to this as the \emph{discriminative loss}. Given that the network architecture comprises bidirectional layers, the max operation in the objective ignores \emph{when} the network produces a large output, since the output at \emph{every time-step} is conditioned on the entire input sequence. The negative loss on the other hand encourages the network to produce a large value for the negative label at \emph{every} time-step.

\subsection{Inference} 

Although the 2 branches of the network (phonetic and discriminative) are trained jointly with all the weights in the biLSTM layers shared, they learn to perform very different tasks. The phonetic branch learns to assign high probabilities to the correct sequence of output labels. During inference, the phonetic branch can be used to compute probabilities for a given phone sequence, e.g. $p(\text{VoiceTriggerPhoneSeq }|\text{ audio})$. Although useful as demonstrated in previous work \cite{sigtia2020mtl,adya2020hybrid}, the phonetic branch cannot be used to score the payload, where we are not sure what the phonetic content in the audio will be. 

The discriminative branch on other hand is trained to perform sequence classification. There are only 2 output classes and there is no concept of a phone sequence. During training, the network must learn which cues in the audio are useful predictors of the correct label. This design is compatible with the task at hand where we do not know what follows the trigger phrase, but we hope that as we input more audio to the network it is able to more accurately predict the correct binary label. Therefore in all experiments presented in this paper, we use the discriminative branch of the \emph{same model} for inference. The phonetic branch can be regarded as a \emph{regulariser} since it is only present during training. 


\subsection{Training Data}
\label{sec:pagestyle}

We use exactly the same phonetic training data as described in \cite{sigtia2020mtl,adya2020hybrid}. This dataset comprises 8000 hours of transcribed audio obtained from intended invocations of the voice assistant. We start with a clean dataset of near-field examples recorded on mobile phones. We then reverberate the clean dataset by mixing with a set of 3000 room impulse responses. Finally, we mix this dataset with echo residuals to simulate the effects of echo cancellation algorithms running on the device \cite{MLBlogFrontEnd}.

The discriminative dataset on the other hand is significantly smaller. The positive set contains 140,000 examples while the negative set contains 40,000 utterances. In previous experiments \cite{sigtia2020mtl,adya2020hybrid}, we only used audio segments that correspond to the trigger phrase. However in this study, we are also interested in making use of the audio \emph{following} the trigger phrase. Therefore, for each example in the training set, we form the following segments: trigger phrase, trigger phrase + 0.5 seconds, trigger phrase + 1 second, trigger phrase + 1.5 seconds, trigger phrase + 2 seconds and the whole utterance. Given that the training set is relatively small, it is possible that the network could overfit to spurious acoustic cues in the training data. By showing the network multiple \emph{views} of the same utterance, we hope to reduce the chances of overfitting.  

\subsection{Model Training}

We use large-batch synchronous stochastic gradient descent \cite{chen2016revisiting} for training the models. Each mini-batch per GPU comprises 128 training examples and we use 32 GPUs in parallel. We use the Adam optimiser for weight updates and an initial learning rate of 0.0008. We use gradient clipping to avoid gradient explosion in the early stages of training, clipping the norm of the gradient to a value of 20. 

\section{Experiments}

\begin{figure}[t]
\begin{minipage}[]{1.0\linewidth}
  \centering
  \centerline{\includegraphics[width=0.8\textwidth]{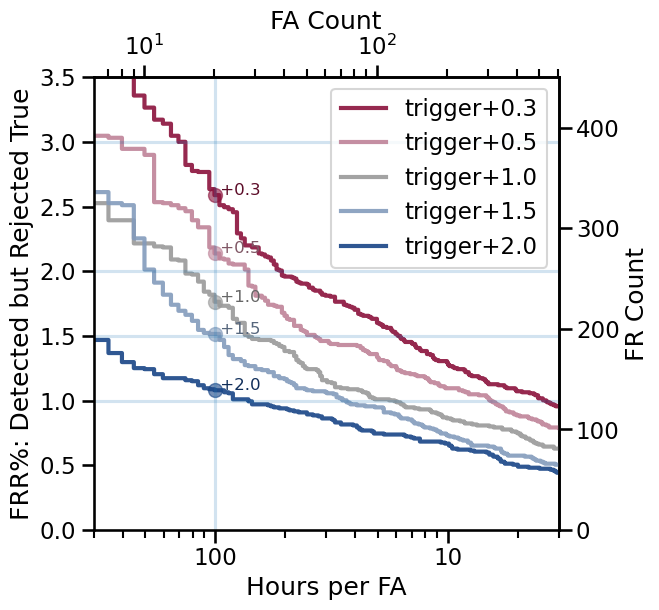}}
  \caption{DET curves as a function of post-trigger audio context. Secondary axes labels show False Alarm (FA) and False Reject (FR) counts.}
  \label{post_trigger}
\end{minipage}
\end{figure}

We use the same evaluation dataset as described in \cite{sigtia2020mtl,adya2020hybrid,sigtia2020mtlspk} without any changes. The test set is the result of a structured data collection where subjects directed a series of voice commands towards the device. There were 100 subjects, approximately balanced between male and female adults. There are over 13k utterances overall, evenly divided between four acoustic conditions: (a) quiet room, (b) external noise from a TV or kitchen appliance in the room, (c) music playback from the recording device at medium volume, and (d) music playback from the recording device at loud volume, the most challenging condition. These examples are used to measure the proportion of false rejections or the False Rejection Rate (FRR). In addition to these recordings, we also use 2,000 hours of continuous audio recordings from TV, radio, and podcasts to measure the false-alarm (FA) rate in terms number of hours of playback per FA.

\begin{figure}[t]
\begin{minipage}[]{1.0\linewidth}
  \centering
  \centerline{\includegraphics[width=0.8\textwidth]{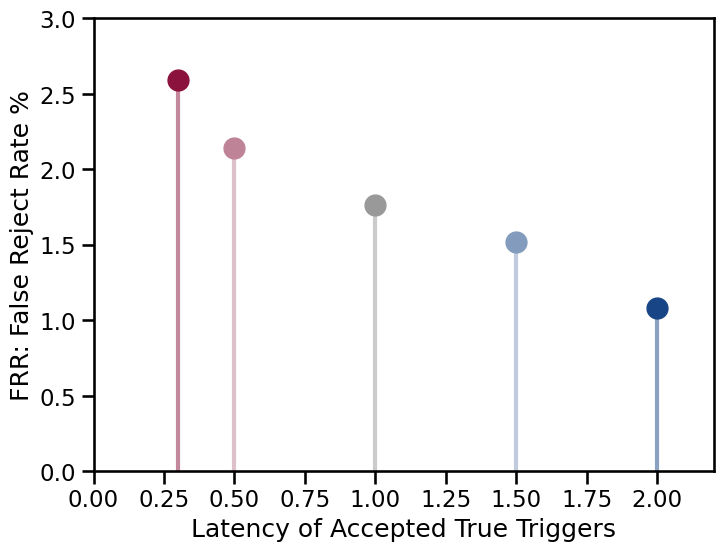}}
  \caption{False Reject Rate at 100 hours per FA as a function of mean latency of accepted true triggers.}
  \label{delay}
\end{minipage}
\end{figure}

As a first experiment, we investigate the effect of adding more audio after the trigger phrase on overall accuracies. As described before, we obtain start and end points for the trigger phrase from a first-pass DNN-HMM model. We then vary the amount of audio added after the end of the purported trigger phrase. We add segments of lengths \{0.3, 0.5, 1, 1.5, 2\} seconds, after the end of trigger phrase. These segments are then input to the discriminative branch of the \emph{same model} described in Section 2 to obtain a score $p(\text{true} | \text{audio\_segment})$. These scores are then used to produce the modified detection error trade-off (DET) curves in Figure \ref{post_trigger}. The x-axis represents number of hours per FA, the y-axis represents the proportion of falsely rejected examples. 
Figure \ref{delay} provides an alternative view where we compare false rejection rate at a fixed point on the the x-axis in Figure \ref{post_trigger}.   
From Figures \ref{post_trigger} and \ref{delay}, it is clear that accuracies do improve as we include more audio context after the trigger phrase. 
For example, by adding 2 seconds of post-trigger audio we are able to more than halve the errors compared to adding only 0.3 seconds of audio. However an additional latency of 2 seconds for every trigger is unacceptable for a practical on-device system. These results suggest that we can get \emph{progressively} more accurate predictions as we listen to more audio. In the extreme case, we could compute scores with every new frame of audio and use a per-frame streaming signal to make decisions. In the next section, we describe a simple two-stage approach where the model produces outputs at only two predefined time-steps.  

\subsection{Two-Stage Design}

\begin{figure}[t!]
\begin{minipage}[]{1.0\linewidth}
  \centering
  \centerline{\includegraphics[width=0.8\textwidth]{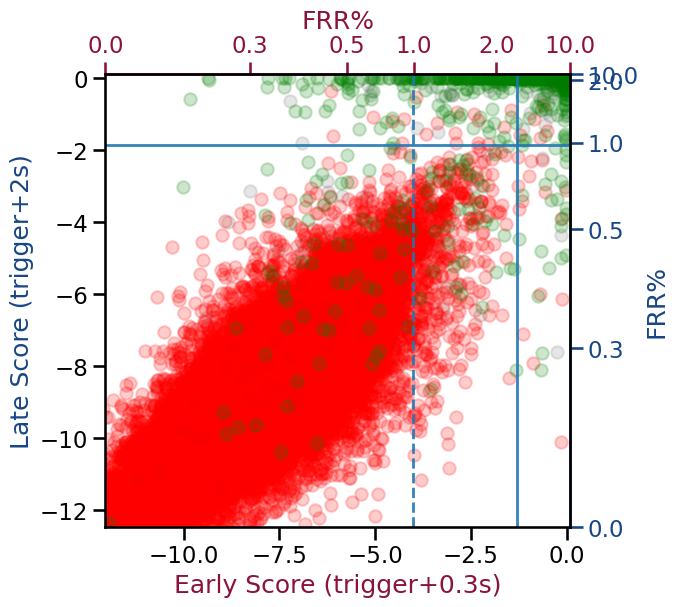}}
  \caption{Scatter plot of early and late scores. Secondary axes labels show False Rejection Rate (FRR)\%. Green circles represent true triggers, red circles represent false triggers.}
  \label{scatter_plot}
\end{minipage}
\end{figure}

We consider a model that produces scores at two intervals, an early score and a late score (Figure \ref{windows}). We choose 0.3 seconds of post-trigger audio for the early score and 2 seconds for the late score for this paper, though these intervals can be chosen differently for different devices and latency budgets. For a two-stage design, we need to consider the joint distribution of an early score and a later one. This joint distribution is presented in Figure \ref{scatter_plot} as a scatter plot. The x-axis represents the early score, while the y-axis represents the late score for every example in the test set. Green circles represent true examples, while red circles represent negative examples. The idea is to choose a threshold on the early score, and accept all candidate triggers if that early threshold is exceeded. We can represent this as a vertical line on the scatter plot. 

The solid vertical line in Figure \ref{scatter_plot} represents a threshold where 3\% of true triggers are rejected (marked on secondary axes). Therefore for these 3\% of true triggers, we delay making a decision until we can compute a late score, which is 2 seconds after the initial trigger. We also need to decide a threshold on the late score, which is represented by the horizontal line in Figure \ref{scatter_plot}. We set the late score threshold such that only 1\% of true triggers are rejected. From the scatter plot is it is clear that with the two-stage design, we are able to recover all of the true triggers in the top-left quadrant of the figure, which were previously rejected when we used only the early score. In Figure \ref{scatter_plot}, for comparison the dashed vertical represents a threshold on the early score where only 1\% of true triggers are rejected. It is clear that there are more false (red) points to the right of the vertical 1\% line than there are to the right of the 3\% vertical plus above the 1\% horizontal line, although the number of true (green) points is about the same. Therefore the two-stage approach provides a way to recover a large number of false rejections \emph{without} significantly increasing the number of false alarms. 

\begin{figure}[t]
\begin{minipage}[]{1.0\linewidth}
  \centering
  \centerline{\includegraphics[width=0.7\textwidth]{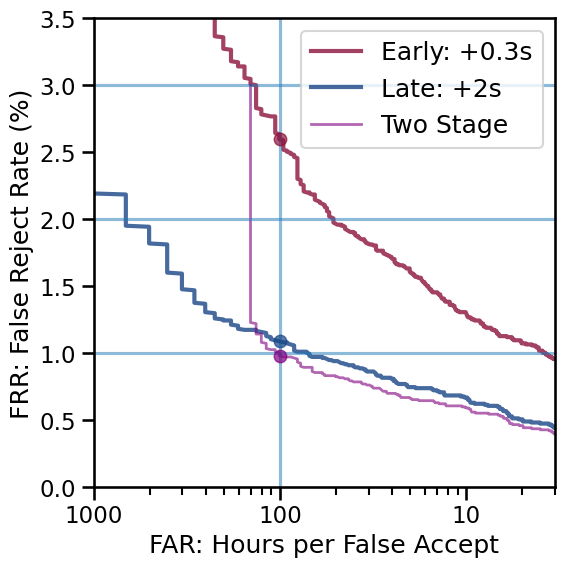}}
  \caption{DET curves for two-stage model.}
  \label{two_stage_det}
\end{minipage}
\end{figure}

Although the scatter plots explain what we propose to do, a more useful way to see the effects on accuracy is in terms of DET curves. In Figure \ref{two_stage_det} we see the DETs for the early and late scores by themselves (red and blue curves respectively). The accuracy of the late curve is clearly better (by a factor of two in regions of interest), but we do not want to wait that long before starting streaming audio. The magenta curve on the other hand shows the accuracy of the proposed two-stage system. It starts on the early curve and departs from it at the chosen FRR of 3\%. Note that two-stage DET curves turns out to be slightly more accurate than using only the late score. 

\begin{figure}[t]
\begin{minipage}[]{1.0\linewidth}
  \centering
  \centerline{\includegraphics[width=0.7\textwidth]{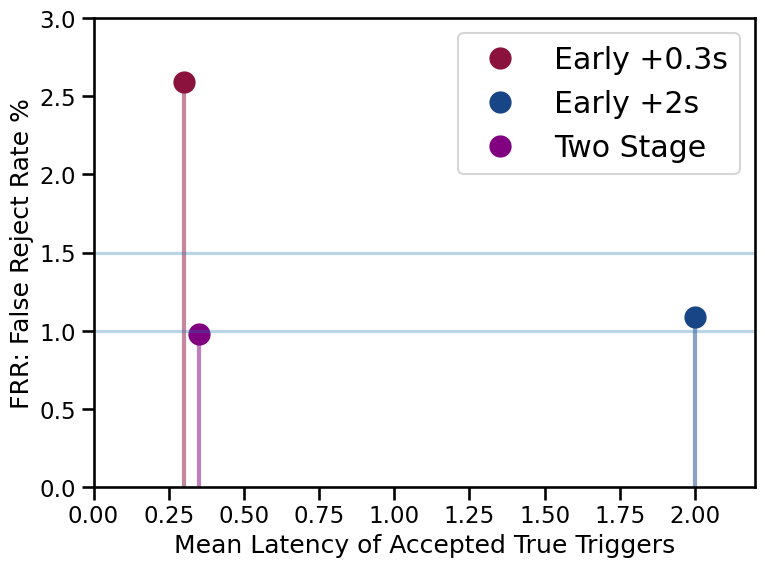}}
  \caption{X-axis plots the mean latency of accepted true triggers in seconds. Y-axis plots the proportion of false rejections. Bar colours correspond to DET curves in Figure \ref{two_stage_det}. }
  \label{latency_plot}
\end{minipage}
\end{figure}

The effect on latency (for true triggers) of the proposed system can be seen be seen in Figure \ref{latency_plot}. Here, we plot the False Rejection rate at 100 hours per false alarm. The mean latency for the early score is 0.3 seconds (red) while for the late score is 2 seconds (blue). The mean latency for the two-stage model is only slightly greater (17\% relative increase) than the early score, while the FR rate is improved by 66\%. 
We listened to a handful of examples that are accepted by the late score but rejected based on the early score (top left quadrant in Figure \ref{scatter_plot}). We found that the late score helps in the following cases: a) when there is a lot of background noise present, processing longer segments results in better scores, b) when the user does not articulate the trigger phrase clearly but the payload is clearly directed towards the assistant, c) when the user repeats the trigger phrase after a failed first attempt.

\vspace{-2.5mm}
\section{Conclusions}

We have presented an architecture for voice trigger detection that uses information in the speech/words that immediately follow the trigger phrase. We first show how to train a model that progressively yields better estimates as we add more audio to the end of the trigger phrase. We then presented a two-stage design, where the model produces an early and a late score. Our analysis shows that we are able to obtain a 66\% relative reduction in false rejections by delaying the decision for only 3\% of true examples in the test set. An obvious short-coming of the model presented here is the fact that we need to recompute the representations for \emph{all} time-steps for both early and late segments, due to the fact that we use bidirectional layers. In the future, we would like to design a model that can share computation between the early and late segments. 

\bibliographystyle{IEEEbib}
\bibliography{refs}

\end{document}